\documentclass[8.5pt,twoside,twocolumn]{article}
\usepackage[super,sort&compress,comma]{natbib} 
\usepackage{mhchem}
\usepackage{times,mathptmx}
\usepackage{sectsty}
\usepackage{balance} 

\usepackage{graphicx} 
\usepackage{lastpage}
\usepackage[format=plain,justification=raggedright,singlelinecheck=false,font=small,labelfont=bf,labelsep=space]{caption} 
\usepackage{fancyhdr}
\usepackage{xcolor}

\begin{document}

\thispagestyle{plain}
\fancypagestyle{plain}{
\fancyhead[L]{\includegraphics[height=8pt]{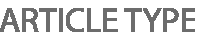}}
\fancyhead[C]{\hspace{-1cm}\includegraphics[height=20pt]{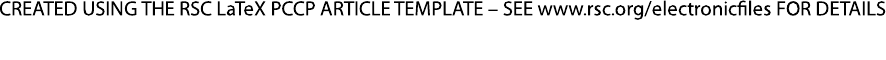}}
\fancyhead[R]{\includegraphics[height=10pt]{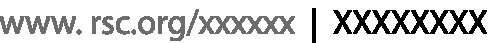}\vspace{-0.2cm}}
\renewcommand{\headrulewidth}{1pt}}
\renewcommand{\thefootnote}{\fnsymbol{footnote}}
\renewcommand\footnoterule{\vspace*{1pt}%
\hrule width 3.4in height 0.4pt \vspace*{5pt}} 
\setcounter{secnumdepth}{5}

\makeatletter 
\def\subsubsection{\@startsection{subsubsection}{3}{10pt}{-1.25ex plus -1ex minus -.1ex}{0ex plus 0ex}{\normalsize\bf}} 
\def\paragraph{\@startsection{paragraph}{4}{10pt}{-1.25ex plus -1ex minus -.1ex}{0ex plus 0ex}{\normalsize\textit}} 
\renewcommand\@biblabel[1]{#1}            
\renewcommand\@makefntext[1]%
{\noindent\makebox[0pt][r]{\@thefnmark\,}#1}
\makeatother 
\renewcommand{\figurename}{\small{Fig.}~}
\sectionfont{\large}
\subsectionfont{\normalsize} 

\fancyfoot{}
\fancyfoot[LO,RE]{\vspace{-7pt}\includegraphics[height=9pt]{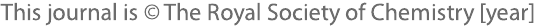}}
\fancyfoot[CO]{\vspace{-7.2pt}\hspace{12.2cm}\includegraphics{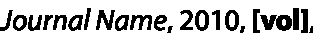}}
\fancyfoot[CE]{\vspace{-7.5pt}\hspace{-13.5cm}\includegraphics{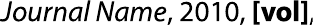}}
\fancyfoot[RO]{\footnotesize{\sffamily{1--\pageref{LastPage} ~\textbar  \hspace{2pt}\thepage}}}
\fancyfoot[LE]{\footnotesize{\sffamily{\thepage~\textbar\hspace{3.45cm} 1--\pageref{LastPage}}}}
\fancyhead{}
\renewcommand{\headrulewidth}{1pt} 
\renewcommand{\footrulewidth}{1pt}
\setlength{\arrayrulewidth}{1pt}
\setlength{\columnsep}{6.5mm}
\setlength\bibsep{1pt}

\twocolumn[
  \begin{@twocolumnfalse}
\noindent\LARGE{\textbf{Root-Growth of Boron Nitride Nanotubes: Experiments and \textit{Ab Initio} Simulations$^\dag$}}
\vspace{0.6cm}

\noindent\large{\textbf{Biswajit Santra,$^{\ast}$\textit{$^{a}$} 
Hsin-Yu Ko,\textit{$^{a}$}
Yao-Wen Yeh,\textit{$^{b}$}
Fausto Martelli,\textit{$^{a}$}
Igor Kaganovich,\textit{$^{b}$}
Yevgeny Raitses\textit{$^{b}$} and
Roberto Car$^{\ast}$\textit{$^{a}$} 
}}\vspace{0.5cm}

\noindent\textit{\small{\textbf{Received Xth XXXXXXXXXX 20XX, Accepted Xth XXXXXXXXX 20XX\newline
First published on the web Xth XXXXXXXXXX 200X}}}

\noindent \textbf{\small{DOI: 10.1039/b000000x}}
\vspace{0.6cm}

\noindent \normalsize{We have synthesized boron nitride nanotubes (BNNTs) in an arc in presence of boron and nitrogen species. We find that BNNTs are often attached to large nanoparticles, suggesting that root-growth is a likely mechanism for their formation. Moreover, the tube-end nanoparticles are composed of boron, without transition metals, indicating that transition metals are not necessary for the arc synthesis of BNNTs. To gain further insight into this process we have studied key mechanisms for root growth of BNNTs on the surface of a liquid boron droplet by \textit{ab initio} molecular dynamics simulations. We find that nitrogen atoms reside predominantly on the droplet surface where they organize to form boron nitride islands below 2400 K. To minimize contact with the liquid particle underneath, the islands assume non-planar configurations that are likely precursors for the thermal nucleation of cap structures. Once formed, the caps are stable and can easily incorporate nitrogen and boron atoms at their base, resulting in further growth. Our simulations support the root-growth mechanism of BNNTs and provide comprehensive evidence of the active role played by liquid boron.}
\vspace{0.5cm}
 \end{@twocolumnfalse}
]

\footnotetext{\textit{$^{a}$Department of Chemistry, Princeton University, Princeton, NJ 08544, USA. Fax: XX XXXX XXXX; Tel: XX XXXX XXXX; 
E-mail: bsantra@princeton.edu, rcar@princeton.edu}}
\footnotetext{\textit{$^{b}$Princeton Plasma Physics Laboratory, Princeton, NJ 08543, USA.}}



\footnotetext{\dag~Electronic Supplementary Information (ESI) available. See DOI: 10.1039/b000000x/}


\section{Introduction}
Boron nitride nanotubes (BNNTs) have attracted great attention because of their extraordinary thermal, mechanical, electronic, and optical properties.~\cite{Golberg2010}
However, the applications of BNNTs have been hampered by the limited yield of the synthesized materials. 
Among many synthesis routes, the nitridation of boron or various boron precursors (\textit{e.g.}, boron oxides) is one of the most popular approaches to grow BNNTs due to its potential for scalability.~\cite{Kim2017} 
The nitridation methods employed to date can be broadly classified into two categories depending on the operating temperature of the synthesis.
The class of low-temperature methods includes chemical vapor deposition (CVD) and ball-milling, typically occurring between 700 K and 2000 K, \textit{i.e.}, below the melting temperature of bulk boron ($\sim$2350 K at 1 bar). 
In ball-milling, precursor powders of boron and metal catalysts are annealed in an N$_2$ or ammonia gas atmosphere,~\cite{Chen1999,Yu2007,FitzGerald2003} whereas in CVD, a boron-containing vapor (for example B$_2$O$_2$), reacts with a nitrogen-containing gas.~\cite{Pakdel2012,Ahmad2015} 
These techniques typically produce large diameter (20--100 nm) BNNTs, and the tubes are frequently found with defects.~\cite{Yu2007,Pakdel2012}
In the class of high-temperature methods boron or boron nitride (BN) powder is vaporized, and the vapor reacts with nitrogen to form BN nanostructures. 
The required temperature can be provided by laser irradiation,~\cite{Lee2001,Arenal2007,Smith2009} arc discharge \cite{Chopra1995,Loiseau1996,Cumings2000,Yeh2017,Kuno2001,Altoe2003} or can be generated in plasma reactors.~\cite{Fathalizadeh2014,Kim2014,Kim2018} 
The tubes produced in the high-temperature method are highly crystalline with few walls (1--5) and small diameters (2--6 nm). 
Early laser irradiation and arc discharge techniques have produced only small quantities (milligrams) of BN nanomaterials due to the difficulty of controlling the nitridation reaction.~\cite{Lee2001,Chopra1995,Loiseau1996} 
In recent years, it was found that radio frequency plasma torches allow to better control thermal and chemical conditions to achieve large-scale production of BNNTs, \textit{i.e.}, several grams per hour.~\cite{Fathalizadeh2014,Kim2014,Kim2018} 
However, in spite of the significant progress in the yield, the growth mechanisms are still poorly understood.

BNNTs are structurally similar to carbon nanotubes (CNTs) suggesting important similarities in the respective growth mechanisms. The growth of the CNTs has been studied extensively with the aim of controlling their chirality, which is a key feature for electronics application. Compelling evidence for a root-growth of CNTs came from in situ measurements using environmental transmission electron microscopy (TEM) in nickel-catalyzed~\cite{Hofmann2007} and in iron-catalyzed~\cite{Yoshida2008} CVD synthesis. In the iron-catalyzed synthesis at 600 $^\circ$C, time-resolved images showed the nucleation of a cap on the surface of iron-carbide nanoparticles and subsequent growth of CNTs \textit{via} root-feeding from the nanoparticle underneath.~\cite{Yoshida2008} Moreover, further support for the root-growth mechanism was provided by atomistic simulations, which suggested that the growth process of CNTs on metal nanoparticles can be decomposed into the following key steps: (i) incorporation of carbon species into the metal clusters, (ii) precipitation of carbon atoms on the surface of the clusters, (iii) cap nucleation, and (iv) growth \textit{via} root-feeding.~\cite{Page2015,Gavillet2001,Raty2005,Ohta2009,Xu2015,Khalilov2015,Neyts2011a}

In contrast to CNTs, much less is known about the growth mechanism of BNNTs. 
A comprehensive effort to understand the growth was provided by Arenal \textit{et al.}~\cite{Arenal2007} In their study, hexagonal-BN ($h$-BN) powder was vaporized by heating with a laser in an N$_2$ environment. 
80\% of the BNNTs produced in this way were found to be single-walled, either isolated or in small bundles.~\cite{Arenal2007} 
Therein, post-synthesis high-resolution TEM images showed that boron particles were attached at the end of the BNNTs, analogously to the images of CNTs attached to metallic nanoparticles.~\cite{Hofmann2007,Yoshida2008} 
Arenal~\textit{et al.}~\cite{Arenal2007} suggested that the growth involves the following steps: (1) vaporization of boron-containing precursors, e.g., boron or $h$-BN powder; (2) condensation of vapor into liquid boron droplets upon cooling; (3) interaction of the droplets with nitrogen-containing species leading to formation of BN caps on the surface of the droplets; (4) nanotube growth \textit{via} progressive incorporation of nitrogen and boron at the interface between the cap and the liquid droplet. 
The root-growth mechanism was also supposed to be effective in other high-temperature BNNT syntheses. 
For example, Smith \textit{et al.}~\cite{Smith2009} were able to increase the BNNT yield by cooling the boron vapor rapidly and flowing N$_2$ gas at high-pressure (2--20 times atm. pressure), which allowed to increase the reaction rate between the boron droplets and N$_2$ molecules. 
In another study, Zettl and co-workers also used high-pressure (up to 10 atm.) N$_2$ gas interacting with condensed boron to produce high-quality BNNTs at an unprecedented rate of 35 gram/hour.~\cite{Fathalizadeh2014}
Essentially, these experiments suggest that boron may play a catalytic role and the interaction of nitrogen-containing species with the boron droplets is a critical step in controlling the yield of BNNTs. 
However, so far no compelling evidence for root-growth mechanism of BNNTs was obtained from \textit{in situ} measurements or atomistic simulations as in the case of CNTs.

The extreme conditions for high-temperature BNNT synthesis are challenging for direct \textit{in situ} imaging. 
Indeed, the \textit{in situ} TEM imaging of CNT growth was performed at temperatures below 900 K,~\cite{Hofmann2007,Yoshida2008} which are much lower than the temperatures required for BNNT synthesis. 
Laser-based in situ diagnostics at high-temperature conditions have just started to emerge,~\cite{Yatom2017,Gerakis2017,Vekselman2018,Yatom2018} but they are yet to be applied to BNNT synthesis. 
A few \textit{in situ} images captured using high-speed cameras provided a qualitative confirmation of the evaporation and condensation process of boron during BNNT synthesis.~\cite{Tiano2014} 
However, knowledge of the conditions required for the organization of BN structures on the boron surface and for the subsequent growth of BNNTs is missing. 
The atomistic details for the root-growth mechanism, \textit{i.e.}, assembly, cap formation, and growth \textit{via} root feeding on the liquid boron surface, are unclear. 
The thermodynamics and kinetics of nitrogen on the surface of boron are fundamental to understand the early stage of BNNT growth. To our knowledge, no atomistic simulations of these non-equilibrium processes have been reported in the literature.

In this work, experiment and theoretical modeling were combined to study the BNNT growth. BNNTs were synthesized by a dc arc discharge in the N$_2$ gas environment at near atmospheric pressure. From post-synthesis high-resolution TEM images, we found that BNNTs were often attached to nanoparticles which are much larger than the diameters of the attached BNNTs. Chemical analysis using energy dispersive X-ray spectroscopy (EDS) shows that the tube-end nanoparticles are composed of boron and do not contain tungsten or any other transition metals, indicating that transition metals are not necessary for the arc synthesis of BNNTs. The TEM images suggest that the boron particles played an important role in the nucleation and growth of BNNTs. To gain microscopic understanding of the early stage of nucleation and growth of BNNTs on the liquid boron surface, we have performed density-functional theory (DFT)-based \textit{ab initio} molecular dynamics (MD) simulations. We used periodic slab models to mimic the near-surface region of a large boron particle. We studied the interaction of the slabs with atomic nitrogen, BN, and N$_2$ molecules in the temperature range 1600--4000 K. We found that nitrogen atoms reside predominantly on the surface of liquid boron and diffuse fast. With increasing nitrogen coverage on the surface, at temperatures in the range 2000--2400 K the nitrogen atoms organize with boron and form BN islands consisting of hexagonal building blocks. The BN islands are not flat but assume curved forms that minimize the contact with the liquid surface underneath. Larger curvatures would further reduce the contact region and could drive the nucleation of nanotube caps. This process would involve coordinated switches of several BN bonds and did not occur spontaneously on the time scale of our simulations. However, once formed a cap is stable on the time scale of the simulations and can easily incorporate additional boron and nitrogen atoms at its interface with the liquid boron particle. These results underlie the catalytic role played by the liquid boron particles, strongly supporting a root-growth mechanism.

\section{Results and discussion}

\begin{figure}
\centering
  \includegraphics[width=1.0\linewidth]{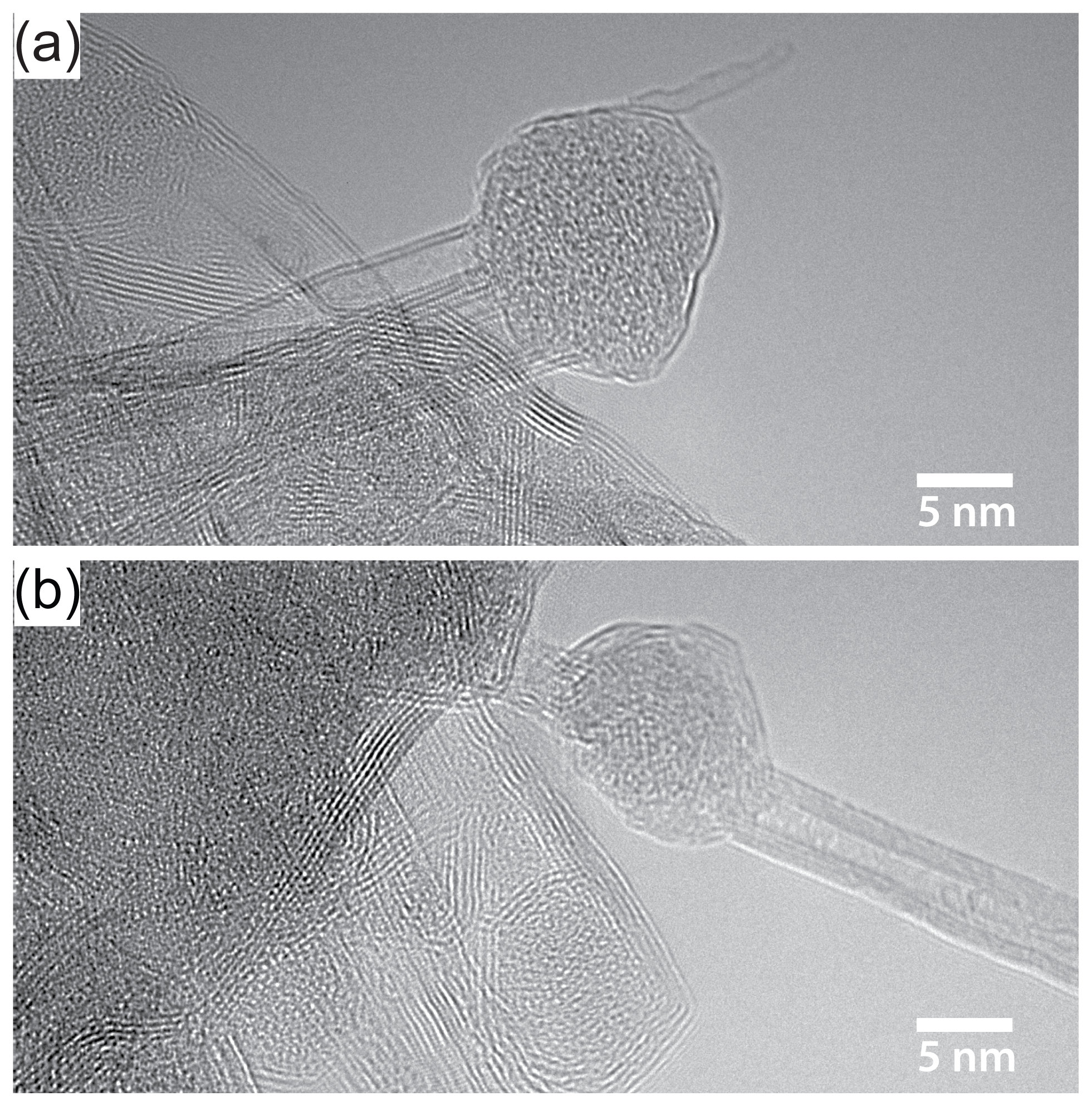}
  \caption{Post-synthesis high-resolution TEM images showing BNNTs attached to a boron nanoparticle. The BN material is obtained from an arc discharge synthesis.}
  \label{fgr:1}
\end{figure}

\subsection{Experimental evidence that BNNTs form on boron droplets in arc} 

In our arc discharge synthesis, a pure boron target was immersed in the arc plasma maintained between two tungsten electrodes at near atmospheric pressure of the N$_2$ gas (Figure S1$\dag$). At the arc temperature, boron vaporizes and then condenses into droplets in the region outside the arc where nanotube synthesis is supposed to occur. The droplets are in liquid form. The post-arc produced BN-containing materials were analyzed using high-resolution TEM. Single and double walled tubes were predominant among all the observed BNNTs. Moreover, the tubes were often attached to nanoparticles. 
Figure \ref{fgr:1} shows two TEM images in which BNNTs are attached at one end to nanoparticles much larger than the radius of the nanotubes. Also, it can be seen from the TEM images that both the nanotubes and the nanoparticles attached at the tube-end have similar brightness. The contrast in Figure \ref{fgr:1} differ from the TEM images reported in earlier arc syntheses of BNNTs.\cite{Chopra1995,Loiseau1996,Yeh2017} In the previous arc studies, the nanoparticles attached at the tube-end had much darker contrast compared to the nanotubes. It was presumed that the darker region of the nanoparticles contained transition metals\cite{Chopra1995} or metal-boride nanocrystals.\cite{Loiseau1996} Recently, chemical analysis using energy dispersive X-ray spectroscopy (EDS) confirmed that the dark contrasts in the arc-synthesized nanoparticles appeared due to the presence of transition metals, e.g., nickel and cobalt,~\cite{Yeh2017} indicating that transition metal may have played some role in the BNNT synthesis using arc. 
However, there is no discernible contrast between the TEM images of the BNNTs and the tube-end nanoparticles as produced in this experiment (Figure \ref{fgr:1}), suggesting that both are composed of similar elements. In order to characterize the chemical contents of the nanoparticles we have performed EDS on multiple samples extracted from our experiment. A typical EDS spectrum is shown in Figure S2$\dag$ of the Supporting Information. The EDS spectrum shows that the nanoparticles are composed of boron and do not contain tungsten or any other transition metals. Our observation indicates that transition metals are not necessary for arc synthesis of BNNTs.

We emphasize that the metal-free growth of BNNTs as observed here is due to the novel arc method employed in our experiment. Previously, the electrodes used for arcing contained $h$-BN or metal-doped boron, and to circumvent the insulating nature of BN, transition metal particles were fused into those electrodes in various ways. For example, Chopra \textit{et al.}\cite{Chopra1995} used anodes made of a BN rod inserted into a hollow tungsten electrode, whereas, Loiseau \textit{et al.} used electrodes made from hot pressed HfB$_2$.~\cite{Loiseau1996} Few other studies were also conducted using electrodes that were made from boron ingots containing $\sim$4\% of transition metal impurities, typically nickel and cobalt.~\cite{Cumings2000,Yeh2017} During arc synthesis, the electrodes were consumed to a large extent and metal particles were typically found in different parts of synthesized materials. In particular, clusters of transition metal particles were found within tube-end nanoparticles, and because of that the role of metals was not clear in arc synthesis of BNNTS.~\cite{Chopra1995,Loiseau1996,Cumings2000,Yeh2017} In contrast, the present work used pure tungsten electrodes and a separate boron source. Due to the higher melting temperature of tungsten the electrodes are consumed much less during our synthesis compared to previous arc.~\cite{Chopra1995,Loiseau1996,Cumings2000,Yeh2017} The combination of pure tungsten electrodes and a feedstock boron-rod containing less than 0.1\% metal impurities are key factors for the metal-free growth of BNNTs in our arc experiment.


As such, BNNTs with radii significantly smaller than the dimension of the nanoparticle to which they are attached constitute a typical outcome of our experiment. These findings suggest that BNNTs form \textit{via} root-based growth on metal-free boron particles. Although BNNTs grown on metal-free boron particles were not observed in earlier arc produced materials, they were observed in laser vaporization~\cite{Lee2001,Arenal2007,Smith2009} and radio frequency plasma torch~\cite{Kim2014,Kim2018} experiments where BNNT synthesis was supposed to be controlled by the root-growth mechanism.

\subsection{Nitrogen in liquid boron} 
To complement the experimental observations and to gain atomistic details of the BNNT growth mechanisms we have performed \textit{ab initio} simulations. Since experiments show that typical boron nanoparticles have dimension much larger than the diameter of the BNNTs we have modeled a boron nanoparticle with a thin periodic slab and focused our attention on the processes occurring near its surface. 
%
%
To connect with the high-temperatures of arc discharge synthesis, we consider a temperature range of 1600 K to 4000 K.
The boron slab exhibits good liquid-like diffusivity for temperatures above 1800 K as illustrated in Figure S3$\dag$. Upon lowering the temperature below 1800 K we observe a sudden drop in the diffusivity indicating onset of glassy behavior.   

\begin{figure*}[ht!]
\centering
  \includegraphics[height=8cm]{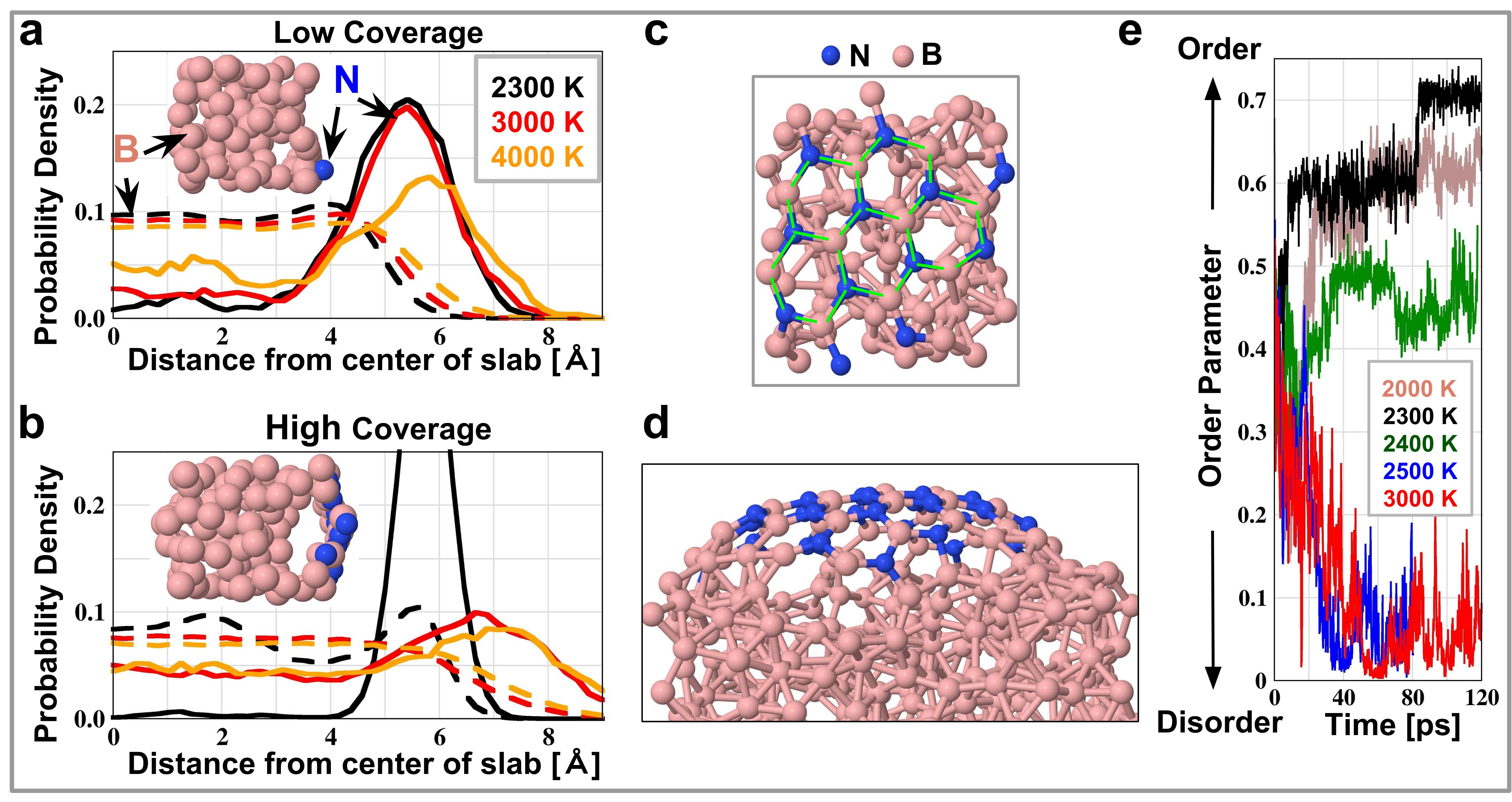}
  \caption{The probability density of the population of boron (dashed lines) and nitrogen atoms (solid lines) as a function of the distance from the center of the liquid slab in the case of (a) the low nitrogen coverage (see text) and (b) the high nitrogen coverage (see text) on the surface of liquid boron. In (b) the inset shows the side view of a representative structure found at 2300 K. (c) The top view of a representative structure found at 2300 K. The BN island consists of several connected hexagonal rings as highlighted in green. (d) The side view of a representative cap-like structure obtained with a larger BN island at 2300 K (see text). (e) The time evolution of the order parameter (see text) in the boron nitride structures at different temperatures.}
  \label{fgr:2}
\end{figure*}

We studied the evolution of nitrogen atoms in the liquid boron slab as a function of nitrogen coverage and temperature. We randomly distributed nitrogen atoms on the surface of the boron slab and computed the probability density of finding the nitrogen atoms at different depths within the slab. We started with a low nitrogen coverage (\textit{i.e.}, one nitrogen atom per 9.1$\times$9.1 \AA$^2$, which is the surface cell of our periodic slab) and accumulated statistics from MD simulations lasting $\sim$250 picoseconds at each temperature. Figure \ref{fgr:2}a shows that at low coverage a nitrogen atom predominantly resides on top of the boron surface, \textit{i.e.}, near the tail of the boron density profile, independently of the temperature. The nitrogen atoms only transiently move to the subsurface layers at 2300--3000 K. At 4000 K, the nitrogen population of the subsurface region increases but only slightly. We observe that a nitrogen atom coordinates with 2--3 boron atoms on the surface and with more than three atoms in the inner layers. These findings indicate that the higher coordination geometry in the subsurface region is less favorable for the nitrogen atom. To further support this observation, we perform few simulations in which a nitrogen atom is initially placed near the center of the slab. In all these simulations the nitrogen atom diffuses from the center towards the surface rather quickly. Also, we find that the nitrogen atoms on the surface diffuse rapidly by switching B-N bonds. The surface diffusivity of the nitrogen atoms ranges from 1.3--4.8 \AA$^2$/ps at 2000--4000 K, and follows closely the surface diffusivity of the boron atoms in the same regime (Figure S3$\dag$). Below 1800 K, the diffusivity of nitrogen atoms drops by more than one order of magnitude, indicating that the rates of the processes depending on the mobility of nitrogen atoms, such as BN organization and growth, should be be severely hampered below 1800 K.

\subsection{Formation of BN islands and cap-like structures} 

With high nitrogen coverage (\textit{i.e.}, twelve nitrogen atoms per 9.1$\times$9.1 \AA$^2$) a more significant fraction of nitrogen atoms populate the inner subsurface layers at 3000--4000 K as shown in Figure \ref{fgr:2}b. However, at 2300 K, the nitrogen population is maximal about $\sim$5.5 \AA\ away from the center of the slab, and is negligible in the sub-surface layers, indicating that all nitrogen atoms reside on top of the surface. On the surface, fast diffusion makes possible frequent interactions among the nitrogen atoms, leading to the formation of small BN chains. Further interactions among these chains stabilize hexagonal BN rings, which are the basic building blocks of BN nanomaterials. In our simulations, it took few tens of picoseconds to form stable BN islands. Figure \ref{fgr:2}c shows a snapshot of a BN island formed at 2300 K.

We also observe that a formed BN island tends to minimize the contact with the boron surface. The density profile of boron at 2300 K (Figure \ref{fgr:2}b) shows that the population of boron increases near the location of the nitrogen density maximum and decreases in the region immediately underneath, indicating a separation of the BN island from the surface. Indeed, the center of the BN island moves away from the surface as shown in the inset of Figure \ref{fgr:2}b. The central part of the BN island binds weakly with the boron slab because the nitrogen and boron atoms at the center of the island have nearly saturated B-N chemical bonds with 3-fold coordination in the plane of the island, similar to the atoms in an sp$^2$ hybridized $h$-BN sheet. The BN island remains attached to the surface only through the atoms at its periphery. The detachment of the center of the island from the surface creates a curved shape which could be a precursor for the formation of a BN cap.

To further investigate if a cap-like protrusion could form spontaneously in our simulation we consider a larger (18$\times$18 \AA$^2$) surface cell and implant a flat BN island (containing 25 nitrogen atoms) on the surface of the liquid boron slab. Here we observe that the implanted flat island acquires a curvature on a timescale of few picoseconds at 2000--2300 K. The curved island is bound to the surface of the slab only at its periphery. Figure \ref{fgr:2}d shows the curved structure in which the central atoms are $\sim$3 \AA\ away from the boron slab, and the peripheral atoms are bound to the slab. The separation of the central atoms of the island from the slab did not increase further in a simulation lasting a few tens of picoseconds. A larger separation would require formation of a BN nanotube cap, which has higher curvature than a BN island. Caps are not made only of 6-membered rings of BN bonds like those present in a spontaneously formed island. For example, the armchair-type BN cap that will be considered later in the paper includes 4-, 6-, and 10-membered rings. Only even membered rings occur to avoid N-N or B-B bonds. Transformation of 6-membered rings into 4- and 10-membered rings involves concerted bond switches and is a rare event on the time scale of our simulation. For example, estimates from tight-binding and force-field based simulations of CNT growth on transition metal clusters suggest that timescales of at least 0.4--1.0 nanosecond are necessary for cap nucleation.~\cite{Ohta2009,Xu2015} Even longer times might be needed in the BN case because of the constraint on the even parity of the rings.

\begin{figure*}
\centering
  \includegraphics[height=8cm]{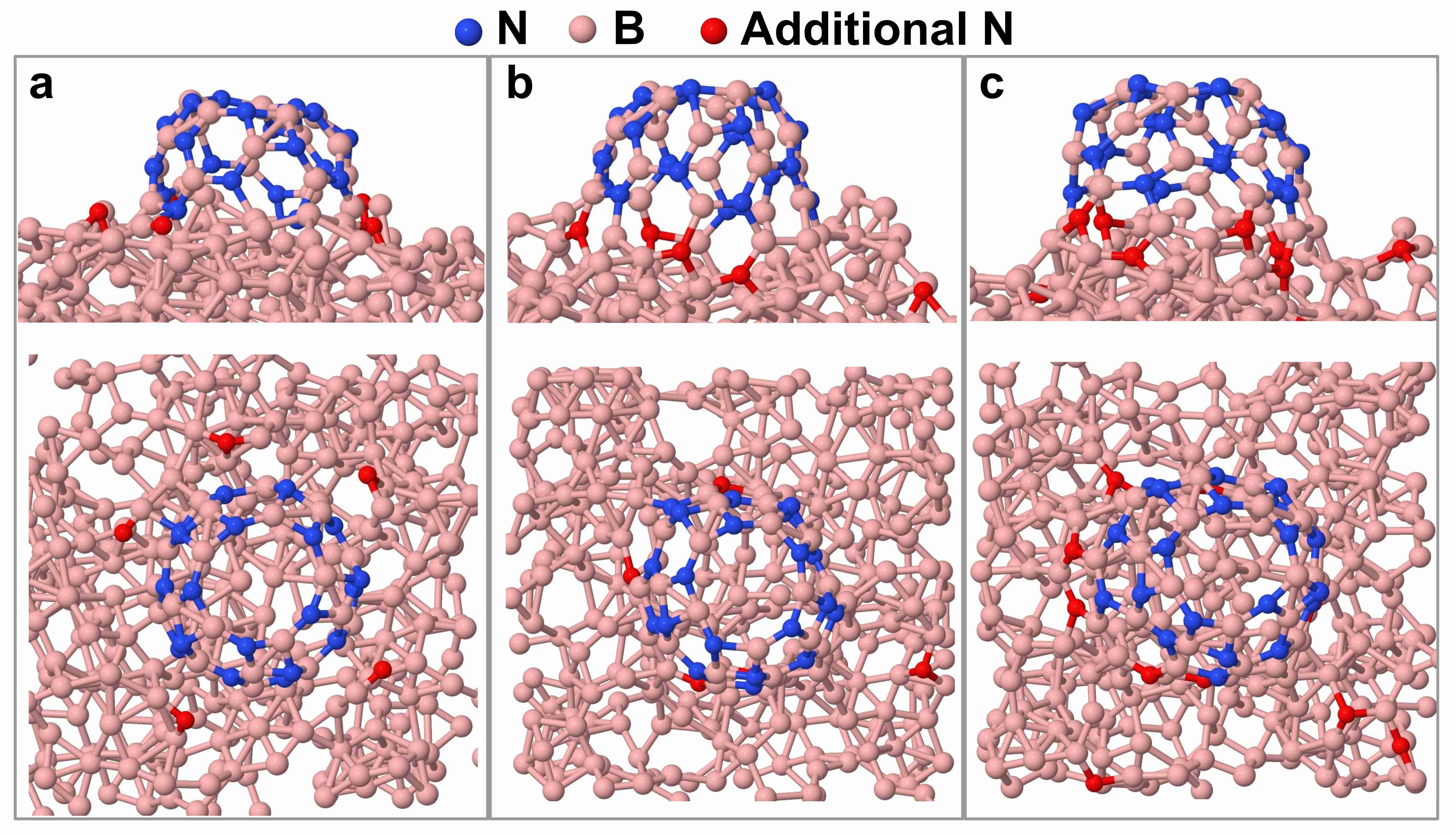}
  \caption{(a) The side and the top view of the initial structure of the BN cap floating on the liquid boron surface with the additional nitrogen atoms. (b) The structure after 5 picoseconds, showing four out of five nitrogen atoms are incorporated at the base of the cap. At this stage, five additional nitrogen atoms are planted on the surface (not shown). (c) The structure after 12 picoseconds, showing six out of ten nitrogen atoms are incorporated at the base of the cap.}
  \label{fgr:3}
\end{figure*}

\subsection{Temperature dependence of BN organization} 
The organization of BN structures on the liquid boron surface strongly depends on temperature. Above 2400 K, no ordered BN islands form on the surface. Instead, we observe the formation of disordered BN mixtures that included atoms in the subsurface layers. A snapshot of a structure obtained at 3000 K is shown in Figure S4$\dag$. To quantify the time evolution of the structural organization of the nitrogen atoms we have utilized a local geometry-matching order parameter, $S$, which measures the overlap between the hexagon formed by the six nitrogen atoms neighboring a nitrogen site and the hexagon formed by the six nitrogen neighbors of a nitrogen site in an ideal planar $h$-BN sheet (section A5$\dag$). The order parameter tends to 1 for perfect matching and tends to zero with increasing disorder. Using this order parameter, we monitor how the system evolves at four different temperatures starting from a configuration in which the nitrogen atoms are randomly distributed on the surface of the boron slab. The evolution of the order parameter is shown in Figure \ref{fgr:2}e. The initial value of $S$ is close to $\sim$0.4, reflecting the initial random distribution. At or below 2400 K, the atoms near the top layer become more ordered as the simulation progresses. We observe formation of hexagonal BN rings that eventually coalesce into a BN island. As a result, the value of the order parameter increases with time. At 2300 K, maximum order ($S \sim 0.7$) is achieved after $\sim$85 picoseconds. At this temperature all the nitrogen atoms belong to the coalesced hexagonal rings as shown in Figure \ref{fgr:2}c. When the temperature is elevated to 2400 K, not all nitrogen atoms belong to hexagonal rings within 120 picoseconds. By further increasing the temperature to 2500--3000 K, the value of the order parameter decreases from its initial value of $\sim$0.4 to values of $\sim$0.1 or less, indicating increasing disorder with several nitrogen atoms in the subsurface layers. 
On the other hand, by lowering the temperature to 2000 K, the BN organization becomes relatively sluggish, and not all nitrogen atoms belong to hexagonal rings within 120 picoseconds. Further reduction in the rate of BN organization is expected to occur at lower temperatures since the surface diffusivity of both nitrogen and boron atoms drops by orders of magnitude for temperatures below 1800 K (Figure S3$\dag$). These findings suggest that the rate of BN organization should be optimal for temperatures in the range 2000--2300 K.

To further investigate the temperature dependent organization of nitrogen atoms solvated in a boron slab, we anneal the system from 3000 K to below 2500 K, starting from an equilibrium structure at 3000 K. Upon cooling to 2000--2300 K, all the solvated nitrogens precipitate from the subsurface layers onto the surface region where they form a well defined island made of BN hexagons only. The solvation and precipitation processes of nitrogen in liquid boron are similar to the more widely studied carbon precipitation in metal nanodroplets in the context of CNT growth.~\cite{Gavillet2001,Raty2005,Ohta2009,Xu2015,Khalilov2015,Neyts2011a} Our findings provide comprehensive evidence that nitrogen atoms segregate from liquid boron and organize into BN islands on the surface when the temperature is below 2400 K.

\subsection{Cap stability and growth \textit{via} root feeding} 
Caps are precursors of nanotube formation. The stability of BN caps on the liquid boron surface is thus critical for the growth of BNNTs. Since we can not observe spontaneous formation of a BN cap in the accessible timescale of our simulation, we construct a cap by thermalizing at 2000 K a cylindrical (5,5) armchair BNNT with an open edge. We observe rapid reconstruction of the open edge into an approximately hemispherical cap to eliminate the dangling bonds present at the open edge. This process leads to formation of 4- and 10-membered rings. We then cut the cap from the BNNT and plant it on top of the liquid boron surface to explore its stability with MD simulations. To perform reasonably long MD simulations, we choose a simulation cell with a sufficiently large surface area (12$\times$12 \AA$^2$) which contains the cap and includes a minimal number of atoms (150) in the overall system (slab+cap). The simulation was run for at least 200 picoseconds at 2000 K. The slab behaves like a good liquid during this entire simulation while the cap diffuses on the surface by switching bonds at the interface between the cap and the liquid surface. More importantly, the floating cap remains upright on the surface and shows no sign of collapsing onto the surface (section A6$\dag$). Our findings suggest that small BN caps are stable on liquid boron surfaces.

According to the root-growth model, a stable cap will further grow upon incorporating boron and nitrogen atoms at its base. Since nitrogen atoms favor to stay on the surface and diffuse on a fast timescale, they are likely to get incorporated at the base of the cap. We perform a few simulations in which nitrogen atoms are fed to the slab surface near a (5,5) BN cap as shown in Figure \ref{fgr:3}a. In this case, we choose a simulation cell with a larger surface area (18$\times$18 \AA$^2$) in order to incorporate additional atoms on the surface. To begin with, we randomly place five nitrogen atoms on the surface of liquid boron as shown in Figure \ref{fgr:3}a. These atoms are placed at least 2.2 \AA\ away from the base of the cap so that they need to diffuse more than one B-B bond length ($\sim$1.7 \AA) to reach the cap. We find that four out of five nitrogen atoms are incorporated at the base of the cap within 5 picoseconds as shown in Figure \ref{fgr:3}b. At this stage, five more nitrogen atoms are randomly placed on the surface at least $\sim$2.2 \AA\ away from the cap. Within 7 picoseconds, two of them are incorporated at the base of the cap as shown in Figure \ref{fgr:3}c. Overall, six out of ten nitrogen atoms are integrated within 12 picoseconds at the base of the BN cap, allowing the cap to grow. We also notice that with higher concentrations of nitrogen on the surface short BN chains formed. The chains diffuse more slowly than isolated atoms and affect the rate of nitrogen incorporation into the cap. This indicates that for sustained growth of the tube, the feeding rate of nitrogen atoms is crucial. We considered a few possible sources for the nitrogen feedstock in high-temperature synthesis. We found that BN, (BN)$_2$, and (BN)$_3$ molecules are easily adsorbed on the liquid boron surface where they dissociate and release atomic nitrogen. On the other hand, N$_2$ molecules interact weakly with the boron slab and do not dissociate if deposited on the liquid boron surface. They do dissociate, however, if they penetrate in the subsurface region (section A7$\dag$).


\section{Conclusions}
In summary, BNNTs were synthesized by a dc arc discharge between two refractory tungsten electrodes with a boron target immersed in the nitrogen arc plasma at near atmospheric pressure. From post-synthesis high-resolution TEM images, we found that BNNTs were often attached to boron nanoparticles. The EDS analysis confirms that the tube-end nanoparticles do not contain tungsten or any other transition metals, indicating that transition metals are not necessary for the arc synthesis of BNNTs. The TEM images indicate that the root-growth is a likely mechanism for BNNTs formation. To better understand this mechanism we have performed DFT-based \textit{ab initio} simulations focusing on the early stages of the BNNT growth. In particular, we studied two key issues in the growth process: (i) the dynamics of the dissolved nitrogen atoms and their organization into BN islands on a molten boron surface and (ii) the stability and growth of a nanotube cap \textit{via} root feeding. We found that atomic nitrogen resides primarily on the surface of a liquid boron droplet, if the temperature is not too high.
%
%
In our simulations, formation of ordered BN islands made of BN hexagons only occurs below approximately 2400 K. Lower temperatures favor formation of ordered structures but for temperatures below 2000 K the rate of island formation slows down and we expect that the entire process should be rapidly quenched as sluggish diffusion sets in below 1800 K. While the precise definition of the optimal temperature regime for growth may depend on the details of the model, such as the adopted DFT approximation, and may be affected by the size and time scale accessible in the simulations, we expect that the general trends observed in our study should be robust.
The islands assume curved shapes to minimize contact with the underlying liquid boron surface and could act as precursors for the nucleation of BN caps having higher curvature. Cap nucleation does not occur on the short timescale of our simulations due to the activation free energy necessary for concerted bond switching. However, once formed BN caps are stable and float on the liquid boron surface over the entire time span of simulations lasting more than 200 picoseconds at 2000 K. At this temperature nitrogen atoms diffuse rapidly on the liquid boron surface allowing the cap to grow. We recall that the melting temperature of bulk boron is 2350 K but that of small droplets is expected to be considerably lower. Indeed the thin slab used in our simulation is still a good liquid at 2000 K. Our simulations support a root-growth mechanism for BNNTs on liquid droplets made of pure boron, consistent with our experimental findings. According to our simulations, the processes leading to cap nucleation and nanotube growth can only occur in the region of the arc chamber where the temperature is about 2400 K or lower. 
%
%
The temperature regime in which BN islands organize on a molten boron surface should also be relevant for other high-temperature synthesis methods, such as \textit{e.g.} induction thermal plasma\cite{Fathalizadeh2014,Kim2014,Kim2018} and laser irradiation\cite{Lee2001,Arenal2007,Smith2009} techniques. The basic root growth mechanisms that we found, such as surface precipitation, island formation, cap nucleation and growth, are broadly relevant to nanotube growth beyond BNNT synthesis and have many common aspects with the mechanisms for carbon nanotube growth on the surface of transition metal particles.~\cite{Page2015,Gavillet2001,Raty2005,Ohta2009,Xu2015,Khalilov2015,Neyts2011a}

\section{Methods}
\subsection{Arc synthesis and electron microscopy} BNNTs were synthesized by a dc arc discharge in a pure nitrogen environment at 400 Torr. The cathode and the anode of the arcs were made from lanthanated tungsten rods of 3.125 mm diameter and 6.35 mm diameter, respectively (Figure S1$\dag$). The arc was generated by briefly bringing the cathode and the anode in contact, after which the current was maintained at 40 A. An external control system increased the electrode gap until the specified discharge voltage (35--40 V) was reached. This voltage includes the voltage drop across the arc and along the electrodes. The gap between the electrodes did not exceed 1 cm wide. A 99.9\% boron target with 0.1\% of metal impurities was immersed into the hot plasma region of the arc discharge. The target evaporated providing boron feedstock for synthesis of BN nanoparticles, including BNNTs. Transmission electron microscopy (TEM) samples were prepared by sonicating the acetone solution that contains the white synthesized product for 2 minutes. The morphology of the synthesized products was studied by using an FEI Talos scanning transmission electron microscope operated at 200 kV. The chemical analysis on the synthesized materials was performed using energy dispersive X-ray spectroscopy (EDS).

\subsection{\textit{Ab initio} simulations} 
All DFT-based molecular dynamics were performed using the Quantum ESPRESSO package.~\cite{Giannozzi2017} We adopted the semi-local approximation of Perdew-Burke-Ernzerhof (PBE~\cite{Perdew1996}) for exchange and correlation in all calculations. Only the valence electrons were treated explicitly, and the interaction of the valence electrons with the nuclei and the frozen core electrons was modeled by norm-conserving pseudopotentials~\cite{Hamann1979} taken from the Qbox public library (http://fpmd.ucdavis.edu/potentials/). A plane-wave kinetic energy cutoff of 40 Ry for the wavefunctions was employed. We performed Born-Oppenheimer molecular dynamics. The electronic wavefunction was minimized at each nuclear step using second order damped dynamics with total energy convergence threshold of 10$^{-5}$ Hartree. Nuclear dynamics was integrated with the Verlet algorithm and a timestep of 2 fs. The ionic temperature was controlled with Nos\'e-Hoover chain thermostats. Using these settings the structure of liquid boron is in good agreement with experiment at 2600 K (section A8$\dag$). The surface of the liquid boron was modeled with a 2D periodic slab. The lattice parameters of the simulation cell were set to reproduce the experimental bulk density (2.3 g/cm$^3$)~\cite{Millot2002} of liquid boron at the melting temperature (2350K) and 1 bar. The smaller supercell used for the 2D slab has a surface area of 9.1$\times$9.1 \AA$^2$ with 100 boron atoms in the cell. With this setting the slab includes approximately 6 layers of boron atoms. The lattice parameter along the surface normal was chosen to be 30 \AA\ so that the periodic images are separated by at least 15 \AA\ of vacuum.  The larger supercell used in the simulations has a surface area of 18$\times$18 \AA$^2$ with a reduced width in the normal direction (3/4th of the width of the smaller unit cell) consisting of 300 boron atoms. The order parameter is defined following the prescription of Ref.~\cite{Martelli2018}. See section A5 in the Supporting Information for more details.

\section*{Acknowledgements}

This work was supported by U.S. Department of Energy, Office of Science, Basic Energy Sciences, Materials Sciences and Engineering Division, a Grant through DOE Contract No. DE-AC02–09CH11466. This research used resources of the National Energy Research Scientific Computing Center (NERSC), which is supported by the Office of Science of the U.S. Department of Energy. Additional computational resources were provided by the Terascale Infrastructure for Groundbreaking Research in Science and Engineering (TIGRESS) High Performance Computing Center and Visualization Laboratory at Princeton University. The authors acknowledge Predrag Krstic, Longtao Han, Shurik Yatom, Vlad Vekselman, Alexander Khrabryi, Mikhail N. Shneider, Bruce Koel, and Rachel Selinsky for fruitful discussions.



\balance


\footnotesize{
\bibliography{rsc} 
\bibliographystyle{rsc} 
}

\end{document}